# Microscopic and thermodynamic evaluation of vesicles shed by erythrocytes at elevated temperatures.

T. Moore[a], I. Sorokulova[a], O. Pustovyy[a], L. Globa[a], D. Pascoe[b], M. Rudisill[b] and Vitaly Vodyanoy[a*]

[a] College of Veterinary Medicine, Auburn University, Auburn AL 36849; [b] College of Education, Auburn University, Auburn AL 36849

(Dated: February 2, 2013)

Erythrocytes and vesicles shed by erythrocytes from human and rat blood were collected and analyzed after temperature was elevated by physical exercise or by exposure to external heat. The images of erythrocytes and vesicles were analyzed by the light microscopy system with spatial resolution of better than 90 nm. The samples were observed in an aqueous environment and required no freezing, dehydration, staining, shadowing, marking or any other manipulation. Temperature elevation, whether passive or through exercise, resulted in significant concentration increase of structurally transformed erythrocytes (echinocytes) and vesicles in blood. At temperature of 37 °C, mean vesicle concentrations and diameters in human and rat blood were $(1.50\pm0.35)\times10^6$ and $(1.4\pm0.2)\times10^6$ vesicles/μL, and $0.365\pm0.065$ and $0.436\pm0.03$ μm, respectively. It was estimated that 80% of all vesicles found in human blood are smaller than 0.4 μm. Thermodynamic analysis of experimental and literature data showed that erythrocyte transformation, vesicle release and other associated processes are driven by entropy with enthalpy-entropy compensation. It is suggested that physical state of hydrated cell membrane is responsible for the compensation. The increase of vesicle number related to elevated temperatures may be indicative of the heat stress level and serve as diagnostic of erythrocyte stability and human performance.

Key words: Erythrocytes, echinocytes, vesicles, entropy, microscopy, heat stress

## 1. Introduction

All organisms in order to maintain wellbeing must preserve a homeostasis, which is constantly affected by internal or external adverse forces termed stressors. Temperature represents one of the most challenging stressors. Although humans have the capability to withstand large variations in environmental temperatures, relatively small increases in internal temperature can lead to injury, heatstroke and even death (Crandall and Gonzalez-Alonso, 2010). Exposure to hot weather is considered one of the most deadly natural hazards in the United States. It was estimated that between 1979 and 2002, heatstroke claimed more American lives than the combined effects of hurricanes, lightning, earthquakes, floods, and tornadoes (Leon and Helwig, 2010).

Heat stress, whether passive (i.e. exposure to elevated environmental temperatures) or via the metabolic heat of exercise, results in pronounced cardiovascular adjustments that might involve the interplay of both local and central reflexes in humans (Crandall and Gonzalez-Alonso, 2010). Central and peripheral fatigue was found to be manifested in reduction of force production that was measured immediately after passive and exercise-Induced heat (Periard et al., 2011). These results agree well with those showing that an elevated core temperature is the primary factor contributing to central fatigue during isometric contractions and hyperthermia (Morrison et al., 2004). Studies with rats have shown that body temperature elevation during exercise is important for induction of exercise-increased shock protein 72 level in rat plasma (Ogura et al., 2008).

Vesicles constitute a heterogenic population of cell-derived microscopic size particles that participate in a wide range of physiological and pathological processes. They derive from different cell types including platelets, red blood cells, leucocytes endothelial, smooth muscle, and cancer cells (Burnier et al., 2009; Castellana et al., 2010; Jayachandran et al., 2008; Orozco and Lewis, 2010; Roos et al., 2010; Rubin et al., 2008; VanWijk et al., 2003; Willekens et al., 1997; Willekens et al., 2003; Willekens et al., 2008; Willekens, 2010). In this work we concentrated on the vesicles in blood plasma, focusing on the red blood cell vesiculation (Greenwalt, 2006) at conditions of elevated temperatures (Araki et al., 1982; Baar, 1974; Bessis, 1972; Chernitskii et al., 1994; Christel and Little, 1984; Gershfeld and Murayama, 1988; Glaser, 1979; Ham et al., 1948; Ivanov et al., 2007; Leonards and Ohki, 1983; Loebl et al., 1973; Longster et al., 1972; Przybylska et al., 2000; Rubin et al., 2008; Samoylov et al., 2005;

Schultze, 1865; Wagner et al., 1986; Walsh and Whitham, 2006).

Mature erythrocytes undergo aging and eventually they are recognized and cleared by macrophages (Rous, 1923; Rous and Robertson, 1917). More recently, it was shown that the aging erythrocyte is characterized by changes in the plasma membrane, that make them selectively recognized by macrophages and then phagocytized in spleen, liver, or bone marrow (Ghashghaeinia et al., 2012). This process is termed eryptosis or programmed erythrocyte death (Bratosin et al., 2001; Föller et al., 2008; Lang and Qadri, 2012; Vittori et al., 2012).

Eryptosis, is characterized by erythrocyte volume reduction, membrane vesiculation (release of small vesicles) and cell membrane phosphatidylserine externalisation. Phosphatidylserine-exposing erythrocytes are recognized by macrophages, which engulf and ingest the dying cells. The regulated form of eryptosis is induced by $Ca^{2+}$ influx, and prevented by cysteine protease inhibitors (Bratosin et al., 2001). Prominent among known triggers of eryptosis (Bosman et al., 2011; Lang et al., 2010; Lutz et al., 1977; Richards et al., 2007) is hyperthermia (Foller et al., 2010).

An elevation in cytosolic $Ca^{2+}$ activity, is known to initiate cell membrane vesiculation, cell membrane scrambling and activation of the cysteine endopeptidase calpain, an enzyme degrading the cytoskeleton and thus causing cell membrane blebbing or bulging. $Ca^{2+}$ further stimulates $Ca^{2+}$-sensitive $K^+$ channels, the following outflow of $K^+$ hyperpolarizes the cell membrane, which causes outflow of Cl. The cellular loss of KCl with osmotically driven water leads to cell shrinkage and phosphatidylserine externalization of eryptotic cells. This leads to macrophage recognition and ingestion of dying erythrocytes (Lang and Qadri, 2012).

It has been demonstrated *in vitro* that the normal red blood cells (discocytes) go through shape changes at variation of temperature. The elevated temperature, induces a series of crenated shapes (echinocytes), characterized by convex rounded protrusions or spicules. Gradually, the spicules become smaller and more numerous and eventually bud off irreversibly, forming extracellular vesicles composed of plasma membrane materials (Christel and Little, 1984; Gedde and Huestis, 1997; Glaser, 1979; Glaser and Donath, 1992; Henszen et al., 1997; Muller et al., 1986). We hypothesize that elevation of temperature during a heat stress may cause transformation of some of the normal erythrocytes (discocytes) into speculated echinocytes that shed membrane vesicles. The increase of the number of membrane vesicles produced during heat stress is relational to a number of discocyte-echinocyte transitions. Therefore, the increase of vesicle numbers following heat stress may be indicative of and proportional to the stress level.

## 2. Methods

### 2.1 Subjects and protocols

#### 2.1.1 Human case study

One healthy male volunteer (age 61) was recruited. One week prior to beginning of heat protocol the volunteer was invited for assessment of maximal oxygen consumption ($VO_{2max}$) and anthropometric data (body weight, body composition, height, etc.). Once a week the volunteer exercised after overnight fast at euhydrated state on the treadmill at a work rate corresponding to 80% of the $VO_{2max}$. Exercise continued each day until core temperatures was elevated 1, 2, and 3°C above resting. Small blood samples (7 μL) were collected before and after exercise. The subject inserted a rectal thermometer 10cm past the anal sphincter for continual monitoring of core body temperature and places a heart rate (HR) on their chest for continual HR monitoring.

#### 2.1.2 Animals

Adult Sprague-Dawley male rats weighing 250-300 g were used. The animals were housed two in a cage under standardized conditions of temperature (20 ± $1^0$ C), humidity (50 ± 5%) and lighting (12-h day/12-h night) with free access to food and water. Two groups of rats (5 rats in control, and 4 rats in stress group) were studied. These groups are: 1) control (25°C) and 2) stress (45°C, 25 min). Animals from group 2 were exposed to 45°C heat stress, relative humidity 55% for 25 min in a 1610 Caron Environmental Chamber control animals (group 1) were exposed to identical conditions as the heat stressed animals, but at 25°C. Four hours after the stress experiments small samples (7 μL) of blood were taken from each rat for live cells and vesicles and examined with light microscopy. Temperature of each rat was measured with an infrared digital thermographic camera (Flir B360 High-Sensitivity Infrared Thermal Imaging Camera) placed 50 cm from the animal. This camera has a thermal sensitivity of approximately 0.1 $^o$C and a spatial resolution of 320×240 pixels. The temperature was measured remotely inside the ear (Vianna and Carrive, 2005).

## 2.2 High resolution light microscopy of blood

The microscope system produces the highly oblique hollow cone of light (N.A. 1.2-1.4). Coupled with a high aperture microscope objective with iris, this system provides two different regimes of illumination. When the iris is closed so that no direct light enters the objective after passing through the object, only refracted, scattered, or diffracted light goes in the objective. If the iris is open in such a way to allow the direct entrance of light into objective the front lens of the objective is illuminated by the annular light produced by the empty cone of light entering the objective. In this case the mixed illumination is produced that combines the darkfield and oblique hollow cone brightfield illuminations. The cardioid condenser is an integral part of the illumination system so that the system comprises a collimation lenses and a first surface mirror that focus light onto the annular entrance slit of the condenser. As a part of the illumination system, the condenser is pre-aligned and therefore additional alignment is unnecessary. The illumination system is positioned in Olympus BX51 microscope by replacing a regular brightfield condenser. The illumination system is connected with a light source (EXFO120, Photonic Solution) by a liquid light guide. The objective used for this work is infinity corrected objective HCX PL APO 100×/1.40-0.70, oil, iris from Leica. The image is magnified with a zoom intermediate lens (2×-U-CA, Olympus), a homebuilt 40× relay lens, and captured by a Peltier-cooled camera (AxioCam HRc, Zeiss) and Dimension 8200 Dell computer. The microscope is placed onto a vibration isolation platform (TMS, Peabody, MA). Live images were recorded with Sony DXC-33 Video camera and Mac OS X Computer (Vainrub et al., 2006). Test images were calibrated using a Richardson slide (Richardson, 1988). A small droplet (7 μL) of freshly drawn blood were placed on a glass slide and coverslipped and positioned on the microscope stage with oil contacts between condenser and objective lenses. Ten image frames of $72 \times 53.3$ μm$^2$ in each sample were photographed, videotaped, and concentrations of discocytes, echinocytes, and vesicle count and diameter were measured by Image-Pro Plus software (Media Cybernetics), providing high-resolution direct-view optical images in real time. The samples were observed in an aqueous environment and required no freezing, dehydration, staining, shadowing, marking or any other manipulation. At least 20 image frames were analyzed for each experimental condition. Each frame contained at average between 50 and 200 vesicles (depending on conditions).

## 2.3 Thermodynamic calculations of the erythrocyte vesiculation

During a normal lifespan, an erythrocyte reduces its volume and surface area in part by process of vesiculation – loss free vesicles. In steady-state conditions, the number of vesicles is constant because of equilibrium between vesicles generated by erythrocytes and vesicles destructed by Kupffer cell mechanisms (Werre et al., 2004; Willekens et al., 2005).

The apparent rate of vesicle destruction at constant temperature can be described by first-order kinetics

$$\frac{dN}{dt} = -kN \quad (1)$$

where $N$ is the concentration of vesicles being destructed, k represents the first-order rate constant, and t, the time of the thermal destruction. The solution of the equation (Eq.) (1) can be expressed as

$$N/N_0 = e^{-kt} \quad (2)$$

where $N_0$ is the initial concentration of vesicles and $N/N_0$ represents the ratio of the concentration of destructed vesicles to the initial vesicle concentration. This fraction defines the ratio of intact vesicles. Taking the logarithm of both sides of Eq. (2), one obtains

$$\log(N/N_0) = -2.303 \times kt \quad (3)$$

A plot of the left-hand side of Eq. (3) versus time (t) yields an estimate of k from the slope. The Arrhenius equation (Segel, 1976) can be used to express the temperature dependence of the first-order activation kinetics:

$$k = Ae^{-\frac{E_a}{RT}} \quad (4)$$

where $R$ is the universal gas constant, $E_a$ represents the apparent activation energy, and $A$ -- the pre-exponential Arrhenius factor. Taking the logarithm of Eq. (4) yields:

$$\log k = -\frac{E_a}{2.303R} \times \frac{1}{T} + \log A \quad (5)$$

If the logarithm of *k* in Eq. (5) is plotted against the reciprocal of temperature, 1/T, then the slope of this graph yields the activation energy ($E_a$), the thermal activation level of transitions from intact to destroyed vesicle. The rate constant of the vesicle destruction process depends on the thermodynamic activation parameters of this transition state and can be described by the Eyring equation (Eyring et al., 1980):

$$k = \frac{k_b T}{h} e^{-\frac{\Delta G}{RT}} = \frac{k_b T}{h} e^{\frac{\Delta S}{R}} e^{-\frac{\Delta H}{RT}} \quad (6)$$

where $k$ is the rate constant, $\Delta G$ is the standard Gibbs free energy of activation, h is Planck's constant, $\Delta S$ is the standard entropy of activation, $\Delta H$ is the standard enthalpy of activation, $k_b$ is the Boltzmann constant, R is the gas constant, and $T$ is the absolute temperature in Kelvin. Taking the logarithm of both sides of Eq. (6), one obtains:

$$\log\left(\frac{k}{T}\right) = \log\left(\frac{k_b}{h}\right) + \frac{\Delta S}{2.303R} - \frac{\Delta H}{2.303R} \times \frac{1}{T} \quad (7)$$

If the plot of the left-hand side of Eq. (7) versus $1/T$ is linear, one can compute the value of $\Delta H$ from the slope, $\Delta S$ from the ordinate intercept and the Gibbs free energy of activation by the relation:

$$\Delta G = \Delta H - T\Delta S \quad (8)$$

From Eq. 4 we can define the energy of activation as:

$$E_a = -R \frac{\partial \ln k}{\partial \left(\frac{1}{T}\right)} \quad (9)$$

The substitution of $k$ in Eq. (9) with the k equivalent relation in Eq. (6) and differentiation of this new expression with respect to $1/T$ shows that the Arrhenius (Eq. 4) and the Eyring (Eq. 6) expressions can be related as:

$$E_a = \Delta H + RT \quad \text{and} \quad (10)$$

$$A = \left(\frac{e k_b T}{h}\right) e^{\frac{\Delta S}{R}} \quad (11)$$

Taking the logarithm of both sides of Eq.11, one obtains:

$$\log A = \log\left(\frac{ekT}{h}\right) + \frac{\Delta S}{2.303R} \quad (12)$$

suggesting a linear relation between the logarithm of the Arrhenius frequency factor and entropy of activation obtained from the Eyring equation for the transitional state. The first-order kinetic model (Eq. 3) was used to determine the best fits of the data with the apparent rate constant $k$. The Arrhenius and Eyring equations were then applied to the data in order to determine $E_a$, $A$, $\Delta G$, $\Delta H$, and $\Delta S$ of the temperature-dependent viability transitions.

For two temperature points $T_1$ and $T_2$ it follows from Eq. 5 that

$$E_a = \frac{2.303 T_1 T_2 \log\frac{k_2}{k_1}}{T_2 - T_1} \quad (13)$$

$$\log A = \log k_1 + \frac{E_a}{2.303R} \times \frac{1}{T_1} = \log k_2 + \frac{E_a}{2.303R} \times \frac{1}{T_2} \quad (14)$$

Substituting $\log A$ in Eq. 14 from Eq. 12 it follows that entropy

$$\Delta S = 2.303R \left[\log A - \log\frac{k_b}{h} - \log(eT)\right] \quad (15)$$

Enthalpy $\Delta H$ is found from Eq. 6 as

$$\Delta H = E_a - RT \quad (16)$$

Using $\Delta S$ from Eq. 15 and $\Delta H$ from Eq. 16, the Gibbs free energy of activation is calculated from Eq. 8.

## 2.4 Statistical analysis

T-test, linear regression, frequency and cumulative frequency, curve fitting and graph plotting were carried out using Microcal™ Origin version 6.0 (Northhampton, MA) and 2010 Microsoft Excel. Geometric mean ($\mu_g$) and geometric standard deviation ($\sigma_g$) for set of measured vesicle diameters ($d_1, d_2, \ldots, d_n$) were calculated using the following equations (Wikipedia, 2012a):

$$\mu_g = \sqrt[n]{d_1 d_2 \ldots d_n} \quad (17)$$

$$\sigma_g = \exp\left(\sqrt{\frac{\sum_{i=1}^{n}\left(\ln\frac{d_i}{\mu_g}\right)^2}{n}}\right) \quad (18)$$

Surface mean diameters ($d_s$) and volume mean diameters ($d_v$) of vesicles calculated from the count mean diameters ($d$) measured by light microscopy by using the following equations (Kethley et al., 1963):

$$\log d_s = \log d + 2.303 \log^2 \sigma_g \quad (19)$$

$$\log d_v = \log d + 6.908 \log^2 \sigma_g \quad (20)$$

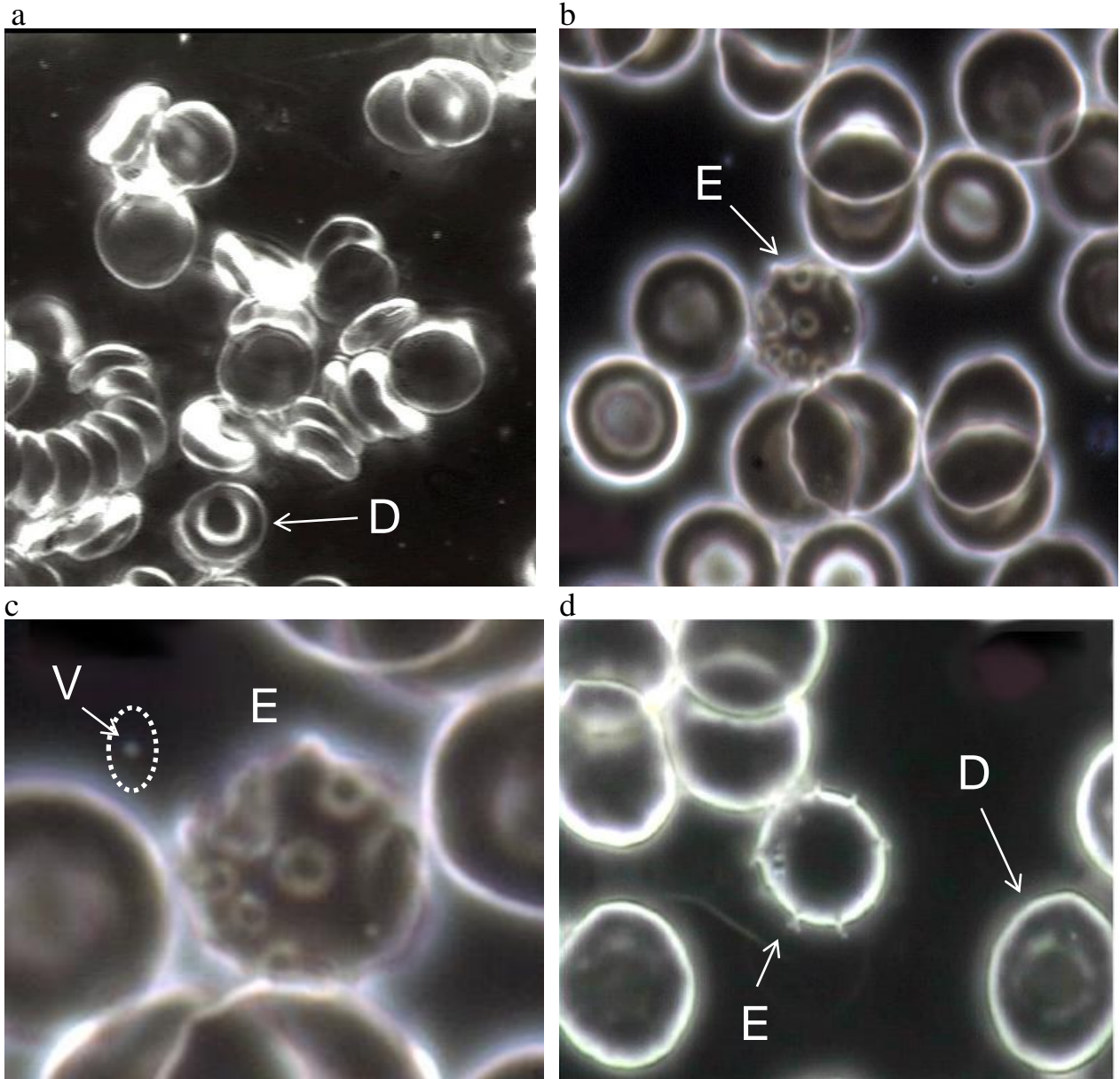

Fig. 1. Human erythrocytes transformation in blood samples at elevated temperatures induced by physical exercise. a. Erythrocytes before exercise at core temperature 37 °C. Most of the erythrocytes are double concave discs of ~7 microns in diameter. D- discocyte (normal biconcave discoid). b. Erythrocytes taken at core temperature of 38 °C after physical exercise. Many erythrocytes are transformed into echinocytes (E) (crenated discs and spheres). c. Enlarged part of the frame b with echinocyte (E) in the middle. A free vesicle (V) is shown in a dashed circle. d. Another picture frame of erythrocytes taken at 38 °C after physical exercise. E – echinocyte (spiculated sphere).

## 3   Results

### 3.1 Erythrocyte-echinocyte transformation in blood

At the basal human core temperature of ~ 37 °C, erythrocytes in freshly drawn blood look as oval and flexible biconcave discocytes with a diameter of ~7 μm ( Fig. 1 a, Table 1).

When temperature rises due to physical exercise, some of the erythrocytes go through size and shape changes, being transformed into echinocytes. Starting with normal discocytes, they become crenated, and then assume spherical shape of smaller diameter with
5

Table 1. Erythrocyte statistics

| Property | Human | | Rat | |
|---|---|---|---|---|
| Life span | 120 days | (Wikipedia, 2012b) | 61 days | (Derelanko, 1987) |
| Diameter | 7 μm | (Evans and Fung, 1972) | 6.9 μm | (Canham et al., 1984) |
| Volume | 90 μm$^3$ | (Evans and Fung, 1972) | 64.9 μm$^3$ | (Canham et al., 1984) |
| Surface area | 136 μm$^2$ | (Evans and Fung, 1972) | 103 μm$^2$ | (Canham et al., 1984) |
| Concentration | 5×10$^6$ μL$^{-1}$ | (Wikipedia, 2012b) | 8×10$^6$ μL$^{-1}$ | (Willekens, 2010) |
| Total number | 2.5×10$^{13}$ (75 kg) | (Wikipedia, 2012b) | 1.3×10$^{11}$ (250 g) | (Lee and Blaufox, 1985) |
| Blood volume | 5 L (75 kg men) | (Wikipedia, 2012b) | 16 mL (250 g rat) | (Lee and Blaufox, 1985) |

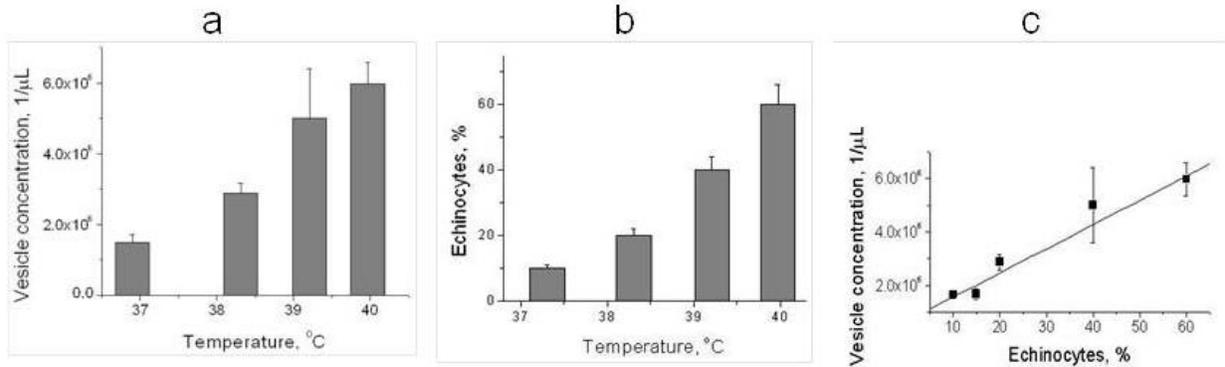

**Fig. 2.** Number of free vesicles is a function of echinocytes population in human blood. **a**. Mean number of vesicles in 1 μl of fresh blood taken before (at ~37 °C, core) and after physical exercise at maximal temperatures of ~ 38, 39, and 40 °C, core, respectively. Count obtained by optical recording of non-fixed 7 μL blood sample with high resolution light microscope system. **b**. Percent of echinocytes estimated in fresh blood taken before (at ~37 °C, core) and after physical exercise at maximal temperatures of ~ 38, 39, and 40 °C, core, respectively. **c**. Linear correlation between number echinocytes and vesicle concentration in blood. Points represent experimental data, while line is a linear fit (R=0.95, P<0.012).

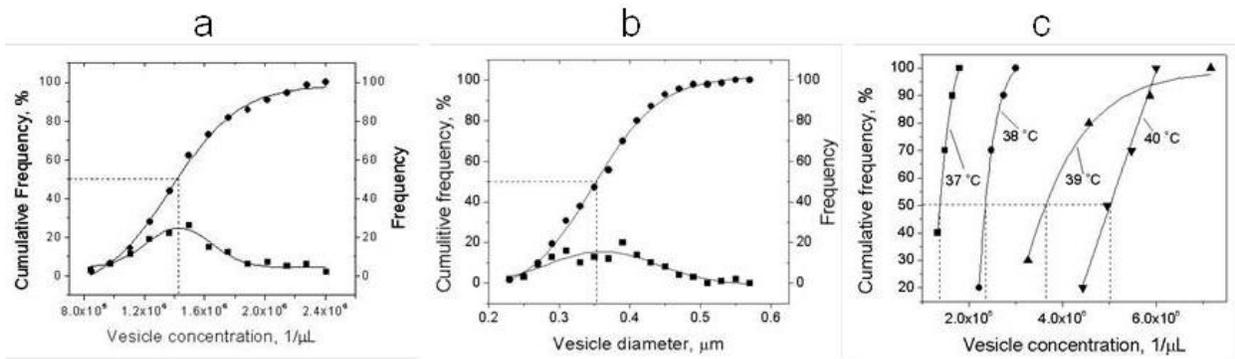

**Fig. 3.** Vesicle count and diameter distribution for human blood. **a**. Vesicle count distributions. Cumulative frequency distribution (●) and frequency count distribution (■) calculated from light microscopy images of non-fixed human blood at 37 °C. The vesicle concentration at the maximum frequency count corresponds the 50% of the cumulative frequency. The points are experimental data while the cumulative frequency and the frequency count curves are the sigmoidal (Boltzmann) and the Gauss fits, respectively. Sigmoidal fit: Chi$^2$ =4.5, R$^2$ =0.997; Gauss fit: Chi$^2$ =4.0, R$^2$ =0.948. **b**. Vesicle diameter distributions. Cumulative frequency distribution (●) and frequency count distribution (■) calculated from light microscopy images of non-fixed human blood at 37 °C. The vesicle concentration at the maximum frequency count corresponds the 50% of the cumulative frequency. The points are experimental data while the cumulative frequency and the frequency count curves are the sigmoidal (Boltzmann) and the Gauss fits, respectively. Sigmoidal fit: Chi$^2$ =4.0, R$^2$ =0.998; Gauss fit: Chi$^2$ =7.7, R$^2$ =0.828. μm. **c**. Cumulative frequency distribution of vesicle concentration as function of temperature. . The data were calculated from light microscopy images of non-fixed human blood at 37, 38, 39, and 40 °C, respectively. The points are experimental data while the cumulative frequency curves are the sigmoidal (Boltzmann) fits. R$^2$ =0.996, 0.987, 0.993, and 0.994 for 37, 38, 39, and 40 °C, respectively.

convex rounded protrusions (buds) (Fig. 1 b and c). The buds grow and finally fall off from the cell surfaces to become free vesicles. Some of the erythrocytes become spherical and grow spicules (Fig. 1 d). Vesicles bud off from the spicules. As temperature increases, the number of echinocytes and free vesicle increases. The images of rat erythrocytes before and after animals exposed to heat are very similar to those of human red blood cells prior to and after physical exercise (data not shown).

### 3.2 Statistics of vesicles shed by red blood cells

#### 3.2.1 Vesiculation of human erythrocytes after exercise

The mean free vesicle concentration in blood after exercise as function of the human core temperature is shown in Fig. 2 a. The concentration is increased approximately 3 times as temperature rises from 37 to 40 $^{o}$C. The percent of echinocytes in blood is also increased with elevated temperature ( Fig. 2 b), so that there is a log-log linear correlation between vesicle concentration and present of erythrocyte-echinocyte transitions.

Fig. 3 a and b show frequency and cumulative frequency distributions of vesicle concentration and diameters at 37 $^{o}$C, respectively. A mean vesicle concentration is $(1.51\pm0.35)\times10^6$ vesicles/µL with a mean diameter (d) is 0.365±0.065 µm. A geometric standard deviation calculated by Eq 19 is equal to 1.20. The surface mean diameter ($d_s$) and volume mean diameters ($d_v$) of vesicles calculated by Eqs 19 and 20 are 0.377 and 0.403 µm, respectively.

Fig. 3 c compares cumulative frequency distribution of vesicle concentration at 37 $^{o}$C (before exercise) and those at temperatures of 38, 39 and 40 $^{o}$C (after exercise). When temperature increases the distribution moves towards higher vesicle concentrations. Vesicle concentration and diameter are shown in comparison with literature data in Table 2.

#### 3.2.2 Vesiculation of rat erythrocytes after exposure to heat (passive hyperthermia)

The life span of human erythrocyte is only ~2 times longer than that of rats (Table 1) in spite of the rat's natural lifespan is about 20 times shorter (Alemáan et al., 1998). Human and rat erythrocyte diameters are very comparable, volume and surface area about 25% large in humans, but concentration is 60% greater in rats. A general appearance of rat blood cells under high resolution microscope is very similar to one observed in human samples (data not shown). As in the case with human erythrocytes, the concentration of free vesicles increase and is followed by the increase in erythrocyte-echinocyte transitions after exposure to heat stress. The concentration of free vesicles increased from $(1.4\pm0.2)\times10^6$ to $(3.8\pm0.3)\times10^6$ vesicles/µL.

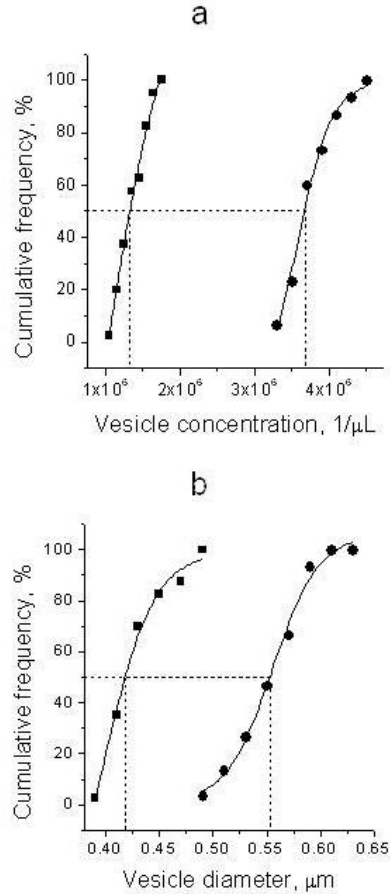

**Fig. 4.** Cumulative frequency distribution of vesicle concentration (**a**) vesicle diameter (**b**) at 36.7 (■) and 43.2 (●) $^{o}$C . The data were calculated from light microscopy images of non-fixed rat blood before and after 25 min at 45 $^{o}$C. The points are experimental data while the cumulative frequency curves are the sigmoidal (Boltzmann) fits. **a**: $R^2$ =0.993 and 0.990, respectively; **b**: 0.991 and 0.994, respectively.

The rat temperature measured in the ear is assumed to be 2.5 degree lower than the rat core temperature (Briese, 1998; Vianna and Carrive, 2005) and increased during a heat exposure from 36.7 to 43.2 $^{o}$C. Fig. 4 shows frequency and cumulative frequency distributions of vesicle concentration and diameter. The figure indicates a significant shift to higher vesicle concentrations and diameter after a heat stress. The mean vesicle diameter at 36.7 $^{o}$C is 0.436±0.03 µm, and at 43.2 $^{o}$C is 0.559±0.03 µm. A geometric standard deviation calculated by Eq 19 is equal to 1.06. At 36.7 $^{o}$C, the surface mean diameter ($d_s$) and volume mean

Table 2. Vesicle concentration (C) and diameter (d)

| Subject | Condition | Method | t, °C | C, µL$^{-1}$ | d, µm | Reference |
|---|---|---|---|---|---|---|
| Human | Live | Light microscopy | 37 | 1.5×10$^6$ | 0.365 | Present work |
| Rat | Live | Light microscopy | 36.7 | 1.4×10$^6$ | 0.436 | Present work |
| Human | Storage, 5 days | Flow cytometry | 4 | 3,370 | <1 | (Rubin et al., 2008) |
| Human | Storage, 50 days | Flow cytometry | 4 | 64,850 | <1 | (Rubin et al., 2008) |
| Human | Fixed | Flow cytometry | 37 | 170 | 0.5 | (Willekens et al., 2008) |
| Human | Centrifuged | Flow cytometry | 37 | 28 | ND | (Berckmans et al., 2001) |
| Human | Centrifuged | Flow cytometry | 37 | 192 | ND | (Shet et al., 2003) |

ND-not determined

Table 3. Thermodynamic analysis of erythrocyte hemolysis and vesiculation

| Process | T °C | Activation energy, $E_a$, KJ.mol$^{-1}$ | Enthalpy $\Delta H$, KJ.mol$^{-1}$ | Entropy $\Delta S$, J.mol$^{-1}$K$^{-1}$ | Entropy contribution T$\Delta S$, KJ.mol$^{-1}$ | Free energy, $\Delta G$, KJ.mol$^{-1}$ |
|---|---|---|---|---|---|---|
| Increase of number of vesicles in blood as result of the human core temperature increase during exercise [1] | 37-39 | 776.1 | 773.5 | 2,342 | 728.3 | 45.2 |
| Increase of vesicles in rat blood as result of environmental heating at 45 °C for 25 minutes[1]. | 36.7-43.2 | 427.4 | 424.8 | 1,240 | 387.6 | 35.1 |
| K+ release from erythrocytes in vitro [2] vesiculation | 19-37 | 63.5 | 66.4 | -19.5 | -24.6 | 91.0 |
| AchE release during vesiculation in vitro [2] | 19-37 | 63.5 | 60.9 | -20.9 | -26.4 | 87.3 |
| Formation of methemoglobin in erythrocytes in vitro [2] | 19-37 | 129.6 | 127.0 | 29.8 | 37.2 | 89.5 |
| Basal ion permeability in erythrocytes in vitro [3] | 37-57 | 102.0 | 99.0 | -5.7 | -7.6 | 106.6 |
| Erythrocyte osmotic fragility [4] | 48-49 | 2400 | 2401 | 7,167 | 2,305 | 96.0 |
| Erythrocyte hemolysis [5] | 38-45 | 131.7 | 129.1 | 71.5 | 22.5 | 106.6 |
| Erythrocyte hemolysis [5] | 45-50 | 342 | 339.1 | 346 | 234.5 | 104.6 |
| Erythrocyte hemolysis [6] | 38-45 | 290 | 286 | 536 | 174 | 112 |
| Discocyte-echinocyte transition [7] | 15-35 | 206.1 | 71.9 | -10.9 | -12.5 | 84.4 |
| Heat survival of CHO cells in cultures[8] | 41-43 | 1266 | 1264 | 3,736 | 1,178 | 86 |

1. Present work. 2. (Chernitskii et al., 1994). 3. (Ivanov et al., 2007) 4. (Ham et al., 1948) 5. (Gershfeld and Murayama, 1988) 6. (Przybylska et al., 2000) 7. (Glaser, 1979) 10. (Bauer and Henle, 1979) diameters ($d_v$) of vesicles calculated by Eqs 19 and 20 are 0.438 and 0.441 µm, respectively.

### 3.3 Thermodynamics of erythrocyte vesiculation

#### 3.3.1 Human case study

When human core temperature rises during exercise from 37 to 38, 37 to 39, and 37 to 40 °C, the change of relative vesicle concentration per minute (apparent rate constant, k) is found to be 8.17×10$^{-3}$, 2.22×10$^{-2}$, and 4.46×10$^{-2}$ min$^{-1}$, respectively. The graph of the logarithm of the rate constants as function of 1000/T (Eq 5) represents the Arrhenius plot shown in Fig. 5 a. The slope of the plot yields the activation energy ($E_a$), the thermal activation level of vesiculation and logA from the ordinate intersect. The same rate constants were used to generate the Eyring plots shown in Fig. 5 b. The linear plot of the left-hand side of Eq 7 vs 1/T brings the value of $\Delta H$ from the slope, $\Delta S$ from the ordinate intercept and the Gibbs free energy of activation ($\Delta G$) by Eq 8. Activation energy ($E_a$) and $\Delta S$ can be calculated by a second way from Eqs 10 and 12. The calculated values of $E_a$, $\Delta H$, $\Delta S$, and $\Delta G$ for process of human erythrocyte vesiculation are shown in Table 3.

#### 3.3.2 Rats

Rat erythrocytes lose 30% of volume during its lifespan (Willekens, 2010). Taking rat erythrocyte volume from Table 1, the lost volume, $L_v = 0.3 \times 64.9$ µm$^3$ =19,47 µm$^3$. The volume of a single vesicle (v) is estimated as $v = \pi d_v^3/6 = \pi(0.441)^3 = 4.49 \times 10^{-2}$ μm$^3$. If all lost erythrocyte volume is converted to vesicles, the number of produced vesicles during the erythrocyte lifespan $n = L/v = 19.47$ μm$^3$/$4.49 \times 10^{-2}$ μm$^3$ = 434.

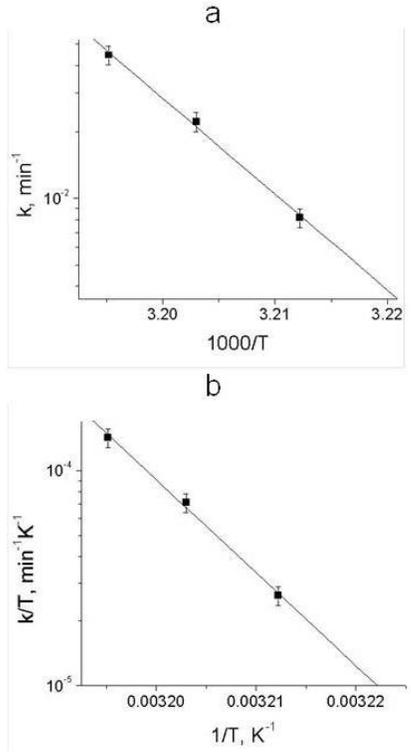

**Fig. 5**. Estimation of vesicle thermodynamic properties by Arrhenius and Eyring equations. **a**. Arrhenius plot. Points are experimental data plotted by Eq 5, line is a linear regression: R=-0.999, P<0.04. **b**. Eyring plot. Points are experimental data plotted by Eq 7 and line is a linear regression. R=-.999, P<0.04.

If the surface area of a single vesicle, $s = \pi d_s^2 = \pi(0.438)^2 = 0.602$ μm$^2$, the total area (S) of 434 vesicles would be equal to $434 \times 0.602$ μm$^2$ = 261 μm$^2$ that is much higher than the whole area of erythrocyte (103 μm$^2$). During a lifespan, erythrocytes reduce their surface area by 20% (Willekens, 2010), which amounts to the area loss $L_s = 0.2 \times 103$ μm$^2$ = 20.6 μm$^2$. Comparing the estimated total area S=261 μm$^2$ to the experimental loss $L_s$= 20.6 μm$^2$, one can conclude that only 20.6/261≈8% of lost area are converted to vesicles. 8% of the lost erythrocyte volume V= 8% × $L_v$=0.8×19.47 μm$^3$=1.54 μm$^3$. Then, number of vesicles produced by a single erythrocyte during its lifespan, $n=V/v$=1.54 μm$^3$/$4.49 \times 10^{-2}$ μm$^3$ =34.22. Taking in account total number of erythrocytes is $1.3 \times 10^{11}$ and lifespan is 61 days (Table 1), it follows that vesicle destruction rate at basal temperature of 36.7 $^o$C, $k_1 = 1.3 \times 10^{11} \times 34.22$:61:24:60≈ $5.0 \times 10^7$ min$^{-1}$.

When the rat is exposed to heat, the rat core temperature increases from $t_1$= 36.7 to $t_2$ = 43.2 $^o$C, vesicle concentration rises from $1.4 \times 10^6$ to $3.8 \times 10^6$ vesicles/μL (Section 3.2.2). Taking in account the duration of hyperthermia (25 min) and rat blood volume (16 mL) (Table 1), the vesicle destruction rate ($k_2$) at the transition from $t_1$ to $t_2$ can be estimated as following

$k_2 = (3.8 \times 10^6 - 1.4 \times 10^6)$ vesicles/μL×$1.6 \times 10^4$μL/25 min=$1.54 \times 10^9$ min$^{-1}$.

Using values of $k_1$ and $k_2$ at absolute temperatures $T_1$ and $T_2$ and Eq. 13, one can estimate the apparent activation energy $E_a$=427.4 kJmol$^{-1}$.

A value of *logA* calculated by Eq 14 and inserted in Eq 15 allows calculation of entropy change (*ΔS*) at temperature T=$(T_1+T_2)/2$ =(309.7+316.2)/2≈313$^o$K (40 $^o$C). *ΔS* is found to be equal to 1240 Jmol$^{-1}$K$^{-1}$.

Inserting values of $E_a$ and RT in Eq 16, one can calculate enthalpy change *ΔH*=424.8 kJmol$^{-1}$.

Finally, substituting *ΔH* and *ΔS* in Eq 8 with found above values, it follows that free Gibbs energy (*ΔG)* is equal to 35.1 kJmol$^{-1}$. The calculated values of $E_a$, *ΔH, ΔS,* and *ΔG* for process of rat erythrocyte vesiculation are shown in Table 3.

### 3.3.3 Thermodynamic correlates of erythrocyte vesiculation

When enthalpy change (*ΔH*) for vesiculation of human and rat erythrocytes obtained in this work together with *ΔH* related to other phenomena related to the thermal cell decomposition given in Table 3 were plotted as function of entropy change (*ΔS),* the data were fitted with a single line that corresponds to the equation *ΔH=ΔG$_o$+T$_o$ΔS*, where *ΔG$_o$*=96.5 kJxmole$^{-1}$ and $T_o$=317 $^o$K (Fig.6).

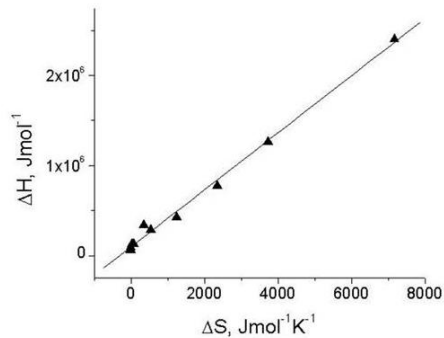

**Fig. 6.** Thermodynamic correlation of enthalpy and entropy for erythrocyte degradation. Points are experimental and literature data (Table 3), while a line is a linear regression: R=0.997, P<0.0001.

# 4 Discussion

## 4.1 Light microscopy of live cells and vesicles

The introduction of no-label high resolution light microscopy gave new opportunity for live cell analysis, including capability of visualizing and recording of cell events at the submicron scale. In the newly developed light system, the object is illuminated by the focused hollow cone of light (instead of beam of light). The effect of the annular illumination promotes the narrowing of the central spot of diffraction pattern and increase of the intensity of the diffraction fringes. The similar effect of annular aperture is known and used in telescopes (Born and Wolf, 1999) to increase the resolution. A better than 90 nm resolution of the Richardson plate (Richardson, 1988) line and picture patterns has been achieved (Foster, 2004; Vainrub et al., 2006; Vodyanoy, 2005; Vodyanoy et al., 2007). The high resolution microscopic illumination system was commercialized and it is known as CytoViva. It is broadly used in live cell and nanoparticle analysis (Asharani et al., 2010; Carlson et al., 2008; Jayachandran et al., 2008; Jun et al., 2012; Lim et al., 2012; Rahimi et al., 2008; Shiekh et al., 2010; Weinkauf and Brehm-Stecher, 2009).

Light images of rat blood cells shown in this work are taken within a few minutes after small samples (<10uL) are drawn. The samples are not subjected any chemical treatments. The cells are videotaped in the natural environment and selective frames are shown in Fig. 1 and 7. The high resolution optical images of human erythrocytes, echinocytes, and free vesicles are consistent with those obtained by electron microscopy (Bessis, 1972; Brecher and Bessis, 1972; Christel and Little, 1984; Dumaswala and Greenwalt, 1984; Longster et al., 1972; Orozco and Lewis, 2010; Repin et al., 2008). The chief advantages of high resolution light microscopy utilized in this work compared to scanning electron microscopy lies in the ability to image live cells in real time at a lower cost. This technique is especially important for free moving vesicles because it allows their discrimination from background.

In this work, vesicle concentrations found in rat and human freshly drawn blood samples at basal conditions found to be at the level of $10^6$ vesicles/μL. This concentration is a few orders of magnitude higher than those found in literature (Table 2). It is well recognized that there is a substantial amount of variation between researchers regarding the preparation procedures used in isolating erythrocyte vesicles, with essential differences in anticoagulants, filtration, centrifugation and flow cytometry, which frequently result in data variability (Dey-Hazra et al., 2010; Hind et al., 2010; Huica et al., 2011). Most of the erythrocyte vesicles are ranged in between 0.1 and 1 μm (Hind et al., 2010), but the most traditional detection technique, flow cytometry, has a detection threshold in the range of 0.4-0.5 μm . The cumulative frequency analysis of vesicle diameters showed that 80% of all vesicles found in blood are smaller than 0.4 μm (Fig. 3 b). it is therefore concluded that majority of vesicles are lost during preparation and detection.

## 4.2 Erythrocyte vesiculation at elevated temperatures

Flow cytometry studies demonstrated that vesicles derive mostly from erythrocytes (approximately 65%) (Willekens, 2010). Human and rat erythrocytes have quite similar physical properties (Table 1). Basal properties including core temperature, vesicle concentration and diameter are very comparable (table 2). When the rat is exposed to environmental hyperthermia, the core temperature and the concentration of free vesicles in the blood increase following the rise of the erythrocyte – echinocyte transitions. This fact is in very good agreement with *in vivo* experiments involving blood and isolated erythrocytes subjected to heat (Araki et al., 1982; Chernitskii et al., 1994; Christel and Little, 1984; Glaser, 1979; Glaser and Donath, 1992; Leonards and Ohki, 1983; Repin et al., 2008; Rumsby et al., 1977; Schultze, 1865; Wagner et al., 1986); and also with erythrocytes involved in burns (Baar and Arrowsmith, 1970; Brown, 1946; Loebl et al., 1973; Vtiurin et al., 1987). It is well established that the increased metabolic heat of physical exercise causes a rise of core temperature in human and animals (Briese, 1998; Greenleaf, 1979; Nybo and Nielsen, 2001a; Parrott et al., 1999; Periard et al., 2012; Sawka and Wenger, 1988; Walsh and Whitham, 2006; White and Cabanac, 1996). In this work, the elevated temperature due to physical exercise in human is accompanied by both the increase in number of transformed erythrocytes (echinocytes) and rise of free vesicle concentration. The thermodynamic analyses of free vesicles in rat and human live blood samples was implemented with the premise that temperature and corresponding energy, enthalpy, and entropy are responsible for red blood cells vesiculation. Obtained thermodynamic data for rat environmental hyperthermia and human exercise induced temperature increase are consistent with each other and agree well with thermodynamic literature results (Chernitskii et al., 1994; Gershfeld and Murayama, 1988; Glaser, 1979; Ham et al., 1948; Ivanov, 2007; Ivanov et al., 2007; Przybylska et al., 2000). Although, all external conditions but temperature were kept the same, the heat stress alone could be suggested as an essential but not the only factor in red blood cell vesiculation. Other stressors may be involved. This situation is similar to another heat stress phenomenon: exercise-increased extracellular heat shock protein 72 (eHsp72) levels. It was shown in rats (Ogura et al., 2008) and human subjects (Whitham et al., 2007), that body temperature increase during physical exercise is essential for activation of exercise-increased eHsp72. However, exercise-related stressors other than heat seemed important in exciting HSP72 release.

A human erythrocyte loses 15 μm$^3$ of volume during its lifespan (120 days) (Bosman et al., 2008). The volume of a single vesicle ($v$) is estimated as $v=\pi d_v^3/6=\pi(0.403)^3=3.43\times10^{-2}$ μm$^3$. If all lost erythrocyte volume ($L$) is converted to vesicles, the number of produced vesicles ($n$) during the erythrocyte lifespan $n=L/v=19.47$ μm$^3/4.49\times10^{-2}$ μm$^3=437$.

If surface area of a single vesicle, $s=\pi d_s^2=\pi(0.377)^2=0.446$ μm$^2$, the total area ($S$) of 437 vesicles would be equal to $437\times0.446$ μm$^2=194$ μm$^2$ that is much higher than the whole area of erythrocyte (136 μm$^2$). During a lifespan, erythrocyte reduces its surface area by 20% (Werre et al., 2004), what amounts to the area loss $L_s=0.2\times136$ μm$^2=27.2$ μm$^2$. Comparing the estimated total area $S=194$ μm$^2$ to the experimental loss $L_s=27.2$ μm$^2$, one can conclude that only $27.2/261\approx14\%$ of lost area are converted to vesicles. Similar estimate made for rat erythrocytes (Section 3.3.2) shows that only 8% of lost erythrocyte material is lost by shedding vesicles. At average, both rat and human red blood cells shed one vesicle every two days during their lifespan. It means that 86% of the lost human erythrocyte material and 92% of rat erythrocyte are lost not by vesiculation but by another unknown mechanism.

A similar analysis of the lifespan loss of human erythrocyte volume and surface area was carried out based on experimental results obtained by centrifugation and flow cytometry (Werre et al., 2004; Willekens et al., 2008). They found a strong discrepancy between estimated and observed loss of erythrocyte surface area. Authors concluded: "Taken together, these data suggest that most vesicles are taken up almost directly by the macrophages of the organ in which they originate before they can reach the venous circulation and be counted. Apparently, the body has developed an efficient mechanism to remove these vesicles..." (Willekens et al., 2008; Willekens, 2010). Being not aware of the direct removal mechanism of hemoglobin by macrophages, the authors explained the extra loss of surface area in comparison to the hemoglobin loss by incorporation of lipids (Bosman et al., 2008). This explanation seems not very likely, because during erythrocyte aging, cell loses 20% of lipids that is equal to a loss of hemoglobin (Rumsby et al., 1977). We hypothesize, that an alternative mechanism of direct transfer of damaged hemoglobin does exist but remains unknown.

During observation and video recording of fresh samples of live blood, direct and prolong contacts between echinocytes and white cells are frequently accounted. In contrast, white cells were not observed in contact with intact erythrocytes. When white cells approach erythrocytes, the cells repulse and scatter. We speculate that during such contacts with echinocytes, peripherally located membrane buds are directly scavenged by white cells. Fig. 7 shows a single frame of a video recording of echinocytes that may be engaged in process of removal of surface vesicles by white cells. The figure also shows also many free vesicles and erythrocytes. The facts of the selective interaction of white cell with echinocytes and the avoidance of erythrocytes are substantiated by the difference in surface electric charges. It has been demonstrated that negative surface electrical charge of aging red blood cells is reduced (Huang et al., 2011). White blood cells have a negative surface charge (Gallin, 1980), therefore the reduction of the aging erythrocyte charge reduces repulsive forces and promotes adhesiveness.

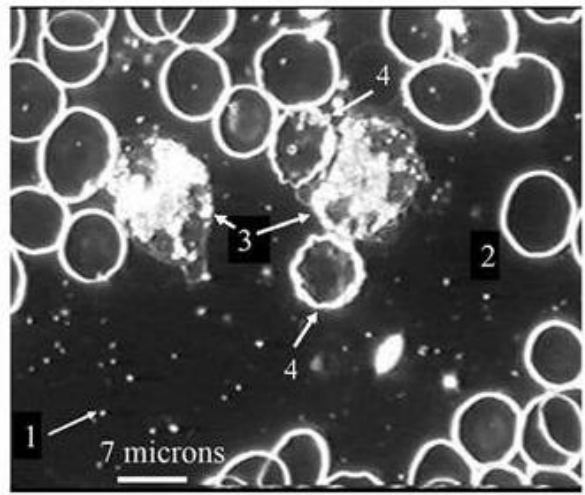

**Fig. 7.** Direct contact of white and red blood cells in human blood. 1-vesicle; 2-erithrocite; 3-neutrophil, 4-echinocytes; 100X objective, oil.

The thermodynamic data include results on rat and human red blood cell vesiculation (obtained in this work), data on individual steps of erythrocyte vesiculation process, and heat survival of Chinese hamster ovary cells (Table 3 and Fig. 6).

It is accepted that intracellular erythrocyte oxidation of hemoglobin into methemoglobin leads to ATP depletion with follow on dephosphorylation of membrane proteins and lipids (Vittori et al., 2012). The change of $K^+$ and $Cl^-$ membrane permeability causes the loss of KCL and osmotically driven water, which in turn leads to cell contraction, externalization of phosphatidylserine (the processes specific to eryptosis) (Lang and Qadri, 2012) and erythrocyte-echinocyte transitions (Backman, 1986). It has been demonstrated that during vesiculation of human erythrocytes in vitro, acetylcholinesterase redistributes on the cell surface and becomes enriched in the released vesicles (Butikofer et al., 1989; Lutz et al., 1977). Additionally, it was shown that acetylcholinesterase (AchE) release correlates with vesiculation of red blood cells (Wagner et al., 1986). It is important to note that process of erythrocyte hemolysis, leads to cell death, but it is very different from the programmed cell death – eryptosis. The main difference lies in the fact that during hemolysis the osmotically driven water moves inside cell, causing it to swell and break it apart, while in eryptosis, water moves outside, causes cell shrinkage and vesiculation. The hemolysis is an adverse cell phenomenon, but eryptosis allows defective erythrocytes to escape hemolysis and prolongs cell life (Föller et al., 2008). When erythrocytes are subjected to hypotonic solution, as in osmotic fragility test (Ham et al., 1948), the cells experience morphological changes that are similar to those during a normal eryptosis. Under influence of osmotic pressure, water moves from inside out, and forces the cells to experience remarkable changes: the first is the appearance of small bud-like protrusions on an occasional erythrocyte; the next recognizable change, many of the erythrocytes show single or multiple buds; then many erythrocytes are converted to spheroid cells, and finally, buds are pinched off creating many free vesicles of one micron and smaller in sizes (Ham et al., 1948).

The change of enthalpy is positive, $\Delta H>0$ for all experiments (Table 3) indicating that all examined processes related to vesiculation and hemolysis are endothermic, i.e. they occur with adsorption of heat. The enthalpy change observed in human erythrocytes due to temperature elevated by exercise is larger than $\Delta H$ in rat exposed to heat. The highest value of $\Delta H$ is observed in the erythrocyte osmotic fragility. It is larger than $\Delta H$ observed in this work. A few conditions of the osmotic fragility experiment are different from those of this work. The osmotic fragility tests are *in vitro* experiments when erythrocytes are isolated and then exposed to hypotonic solutions and elevated temperatures. In experiments of this work, red blood cells are challenged with elevated temperatures while they are *in vivo*. One can speculate that being *in vivo*, erythrocytes may use ATP internal energy to undertake steps leading to change of membrane ion permeability and create osmotic compression. It would save the thermal energy that breaks cells. In contrast, in osmotic fragility experiment, the ATP component is missing and therefore, a higher enthalpy is required to vesiculate cells. A much smaller $\Delta H$ observed in hemolysis of erythrocytes when osmotic pressure drives water inside the cell and swells it. The smaller $\Delta H$ required for cell hemolysis agrees well with the fact that model vesicles can withstand smaller stretching than compressive osmotic pressures (Oglecka et al., 2012). The important step of vesiculation, discocyte-echinocyte transition, has a relatively small $\Delta H$ that is about the same values as enthalpies of $K^+$ and AchE release from erythrocytes during vesiculation.

The values of $T\Delta S$ in Table 3 ranges from -26.4 to 2,305 KJmol$^{-1}$ and for each process they are comparable with the enthalpy change $\Delta H$. Entropies ($\Delta S$) of the ion permeability change and $K^+$-ion and AchE release from erythrocytes during vesiculation are found to be negative. $\Delta S<0$ has conventionally been associated with decreases in molecular rotational and translational degrees of freedom. $\Delta S<0$ in processes involving aqueous solvents often involves the dehydration of solvent moieties forming the interface between the contacting molecules and by release of ions (Dragan et al., 2003). A discocyte-echinocyte transition is also shows a negative entropy change. The appearance of protrusions on the surface of echinocytes, increases the cell membrane curvature and consequently, increases entropy (Beck, 1978). As result, the discocyte-echinocyte transition is accompanied with the negative entropy change.

The energy associated with the change of entropy, $T\Delta S$, correlates with the change of enthalpy, $\Delta H$, keeping the Gibbs free energy, $\Delta G = \Delta H - T\Delta S$, and small compared to both $\Delta H$ and $T\Delta S$. This thermodynamic phenomenon is known as enthalpy-entropy compensation (Leffler, 1955; Tomlinson, 1983)The enthalpy-entropy compensation phenomenon is characteristically displayed by a linear graph of $\Delta H$ vs. $\Delta S$ as revealed in these studies when the experimental data across different results related to red blood cells destruction were grouped in Table 3 and Fig. 6. The $\Delta H$ versus $\Delta S$ results are consistent with the linear free energy relations (r = 0.997, p<0.0001) of a characteristic enthalpy–entropy compensation plot, with the slope approximating the mean temperature of the range of temperatures used in the experiments.

The joint influence of a small temperature range of apparent rates of destruction and the alteration of coordinates can result in entirely mistaken associations between *ΔS* and *ΔH* (Sharp, 2001). In present work, nevertheless, the parameters were determined not separately from pairs of rate data, but obtained by regression analysis of Arrhenius plots that incorporated several rate measurements at the relatively large temperature range of 15-57° C; most of the time the slopes were statistically significant at the 1-4 % level of probability. Since the correlation shown in Fig. 6 is highly significant (P = 0'0001), it is concluded that it is a correct presentation of the compensation law. According to accepted interpretation data like this, this result means that with exception of CHO data, mechanisms of erythrocyte hemolysis and vesiculation is identical for the cells irrespectively of subjects. The suggestion that apparent rates associated with erythrocyte vesiculation, hemolysis, K+ and AchE release, ion permeability, formation of methemoglobin, osmotic fragility, discocyte-echinocyte transition, and CHO heat survival of three different species follow the same mechanism is hard to accept. Nevertheless, all processes include a common substance: they all involve water. Therefore, it is deduced that the *ΔS-ΔH* correlation shown here relies more upon the physical state of water molecules in cell membrane. This explanation is in agreement with the conclusion made in the study of erythrocyte hemolysis of man, rabbits, guinea-pigs, rats, sheep, cats and pigs (Good, 1967) .It is also consistent with deduction of Ives and Marsden , who suggested that for chemical systems the compensation law is most often seen in reactions that vary from each other primarily in the degree to which solvational changes may take place (Ives and Marsden, 1965). More recently, solvent reorganization was suggested as a driving force for encapsulation in water, that can be revealed enthalpy-entropy compensation (Leung et al., 2008). Lipid as a common component of all cell membranes (Yeagle, 1989), is present in all erythrocytes, vesicles, and other cells analyzed in this work. It was demonstrated that hydration highly impacts the molecular and electronic structure of membrane phospholipids and therefore direct the interactions of the lipids with other molecules and strongly contribute to physical properties of membrane (Mashaghi et al., 2012). Thus we speculate that the common mechanism of the *ΔS-ΔH* compensation plot may therefore include the hydration state of membrane phospholipid.

### 4.3 Diagnostic value of erythrocyte vesicles

During physical exercise, body temperature is elevated (Buono et al., 2005; Ogura et al., 2008). One of the serious consequences of a mild increase of temperature (~ 38 °C) is energy depletion (Kozlowski et al., 1985). That is in turn results in the accelerated erythrocyte-echinocyte transition (Backman, 1986; Brecher and Bessis, 1972) and increased vesicle shedding from speculated cells (Christel and Little, 1984). At temperatures higher than 38 °C, a cell's hemolysis comes into effect, and vesicle production grows as temperature increases (Gershfeld and Murayama, 1988). It is therefore, hypothesized that increase of vesicle number related to elevated temperatures during exercise may serve as diagnostic of erythrocyte stability. Because the decreased stability of red blood cells reduces their capability of oxygen transport (Nybo et al., 2001; Nybo et al., 2002; Nybo and Nielsen, 2001b), impairs endurance, and causes fatigue (El-Sayed et al., 2005; Periard et al., 2011), the number of vesicles increases due to exercise may be examined for its diagnostic value of human performance.

## 5 Conclusions

1. High resolution light microscopy of live blood cells is an appropriate technique for imaging heat induced structural erythrocyte transformations and vesicle release.
2. Vesicle release accounts for only a small part of erythrocyte surface reduction during its lifespan. It is suggested that a large part of the aged erythrocyte membrane is removed by direct contacts with white cells.
3. Entropy-enthalpy compensation is associated with erythrocyte-echinocyte transitions at elevated temperatures and is assumed to include the hydration state of membrane lipid.
4. Physical exercise results in elevation of body temperature, energy depletion, acceleration of erythrocyte-echinocyte transition, and in turn increases the number of released vesicles. These changes reduce erythrocyte stability, capability of oxygen transport, impair endurance, and cause fatigue. Therefore, the number of vesicles released increases due to exercise and therefore has a diagnostic value of human performance.

### Acknowledgements
Authors are grateful to members of Department of Anatomy, Physiology and Pharmacology and Department of Kinesiology of Auburn University for support and encouragement.